# Non-equilibrium Green's function based model for dephasing in quantum transport

Roksana Golizadeh-Mojarad and Supriyo Datta
School of Electrical and Computer Engineering, Purdue Universtiy, West Lafayette, IN-47906, USA.
(Dated: December 07, 2006)

The objective of this paper is to describe a simple phenomenological approach for including incoherent dephasing processes in quantum transport models. The presented illustrative numerical results show this model provides the flexibility of adjusting the degree of phase and momentum relaxation independently that is not currently available in mesoscopic physics and in device simulations while retaining the simplicity of other phenomenological models.

*Introduction:* Although models for coherent quantum transport are fairly well established [1], approaches for including incoherent dephasing processes represent an active area of current research. Much of the work is motivated by the basic physics of conductance fluctuation in chaotic cavities [2][3][4]. However, there is also a strong motivation from an applied point of view [5]. For example, if we calculate the transmission $T(E)$ through a 2D conductor (which could be the channel of a nanotransistor) with a random array of scatterers (due to defects or surface roughness), we see fluctuations as a function of energy which arise from quantum interference. Such fluctuations however are seldom observed at room temperature in real devices (when the device length is larger than phase relaxation length), because interference effects are destroyed by dephasing processes. Clearly, realistic quantum transport models for nanotransistors need to include such dephasing processes. A common way of including dephasing is through additional Buttiker probes [6] whose effects can then be modeled within the Landauer-Buttiker framework for coherent transport theory. Such probes typically introduce an additional resistance, since the probes themselves destroy momentum of itinerant electrons by partially reflecting them. By using a pair of unidirectional probes, Buttiker introduced phase relaxation without introducing momentum relaxation [7]. However, we are not aware of any work extending this to a continuous distribution of probes as needed to model a long conductor.

In real devices, dephasing processes often arise from "electron-electron" interaction scattering, which destroy phase but not momentum. The non-equilibrium Green's function (NEGF) method [8][9] provides a rigorous prescription for including any dephasing process to any order starting from a microscopic Hamiltonian through an appropriate choice of the self-energy function $\Sigma_s(E)$ [10][11][12]. The **objective** of this paper is to describe two simple phenomenological choices of $\Sigma_s(E)$ that allows one to incorporate phase relaxation with or without momentum relaxation. In this paper, we restrict our discussion to elastic dephasing where strictly speaking, no energy is exchanged between the electrons and the dephasing source. Real dephasing mechanisms often involve some energy exchange but this has negligible effect on the current if the bias and thermal energy ($eV + k_BT$) are small compared to the energy range $\varepsilon_c$, over which the transmission characteristics remain essentially unchanged (see discussion on P. 105 of Ref. [14]). Our examples largely correspond to this transport regime with large $\varepsilon_c$, which excludes strong localization. Even if $\varepsilon_c$ is small, our elastic dephasing model can be used, but it will not capture effects like hopping that require inelastic interaction. Finally, we note that our purpose is not to provide a microscopic theory of any specific mechanisms (which is already available in the literature [11]). Rather it is to provide an NEGF-based phenomenological model that is comparable to the Buttiker probe model in conceptual and numerical simplicity, and yet allows one the flexibility of adjusting the degree of phase and momentum relaxation independently.

*Theoretical formulation:* The basic theoretical model presented here is based on the general NEGF approach, which can be used to analyze a variety of devices beyond the 1D and 2D geometries explored in this article [13][14][15]. The structure is partitioned into channel and contact regions (Fig. 1) with the channel properties described by a single band effective mass Hamiltonian ($H$). Finite difference approximation is used to find the matrix representation of $H$ assuming uniform lattice spacing '$a$' in all dimensions.

$$H(i,j) = \begin{cases} zt + U(i,j), & i = j \\ -t, & \text{if } i, j \text{ are nearest neighbors} \\ 0, & \text{Otherwise} \end{cases} \quad (1)$$

where $t = \hbar^2/(2m^*a^2)$, $m^*$ is the electron effective mass, $z$ represents the number of nearest neighbors ($z = 2$ for 1D model and $z = 4$ for 2D model), and $U$ stands for the potential profile in the channel. The potential matrix $U$ is

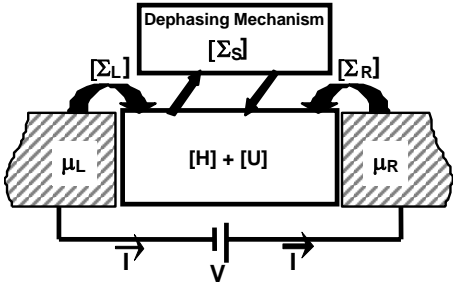

FIG. 1. General schematic illustration used for NEGF quantum transport calculation. Contacts are assumed to remain in local equilibrium with the electrochemical potentials $\mu_L$ and $\mu_R$.

diagonal with the diagonal elements $U(i,i)$ equal to the potential at the lattice site '$i$'. The potential is a sum of two components: one due to defects, impurities, surface roughness, etc. and one due to the applied bias. In this paper, the second component is ignored, as we only discuss problems involving a small bias voltage for which the transmission properties of the structure are approximately constant over the energy range of interest. The effect of contacts on the channel is included through the self-energy matrix $[\Sigma_{L,R}]$ whose elements are given by

$$\Sigma_{L,R} = \tau_{L,R} g_{s(L,R)} \tau_{L,R}^+ \qquad (2)$$

where $\tau$ is the coupling matrix between the contacts and channel. The Hamiltonian for the 2D semi-infinite contacts using Eq. (1) will be in the form of a block tridiagonal matrix with the Hamiltonian matrix ($\alpha$) for an isolated layer of the contacts on the diagonal and the inter-layer coupling matrices $\beta$ and $\beta^+$ on the upper and lower diagonals. The surface Green function for the contacts can be obtained from:

$$g_s = [(EI - \alpha) - \beta g_s \beta^+]^{-1} \qquad (3)$$

where $E$ is energy. This equation can be solved analytically [16] or iteratively starting from a reasonable guess for $g_s$. The channel Green's function is obtained from:

$$G(E) = (EI - H - \Sigma_L - \Sigma_R - \Sigma_s)^{-1} \qquad (4)$$

where $\Sigma_s$ is the self-energy matrix due to dephasing processes. Once $H$, $\Sigma_L$ and $\Sigma_R$ are known, all quantities of interest can be calculated from the following set of equations (the dephasing-related quantities $\Sigma_s$, and $\Sigma_s^{in/out}$ are discussed later):

$$A(E) = i[G(E) - G^+(E)] = G^n(E) + G^p(E) \qquad (5)$$

$$G^{n,p} = G(\Sigma_L^{in,out} + \Sigma_R^{in,out} + \Sigma_s^{in,out})G^+ \qquad (6)$$

$$\Sigma_{L,R}^{in}(E) = f_{L,R}(E)\Gamma_{L,R}(E) \qquad (7)$$

$$\Sigma_{L,R}^{out}(E) = [1 - f_{L,R}(E)]\Gamma_{L,R}(E) \qquad (8)$$

$$\Gamma_{L,R}(E) = i[\Sigma_{L,R}(E) - \Sigma_{L,R}^+(E)] \qquad (9)$$

$$I_i = (e/h)\int_{-\infty}^{+\infty}[Trace(\Sigma_i^{in}A) - Trace(\Gamma_i G^n)]dE \qquad (10)$$

where $G^n/G^p$ ($-iG^</+iG^>$) refers to the electron/hole correlation function whose diagonal elements are the electron/hole density, $A$ is the spectral function whose diagonal elements are the local density of states, $f_{L,R}(E)$ represent the Fermi functions for the related contacts, and $I_i$ is the current at the terminal '$i$'. In our elastic dephasing model, each energy channel is independent so that the current in the channel can be written in the form

$$I_L = (e/h)\int_{-\infty}^{+\infty}\bar{T}_{eff}(E)(f_L(E) - f_R(E))dE \qquad (11)$$

where $\bar{T}_{eff}(E)$ represents the effective transmission of the channel. In general, with dephasing present, there is no direct way of calculating $\bar{T}_{eff}(E)$. Instead, we obtain it by comparing Eq. (10) and Eq. (11) (see P. 291 in Ref. [13]).

$$\bar{T}_{eff} = (e/h)[Trace(\Sigma_L^{in}A) - Trace(\Gamma_L G^n)]/(f_L - f_R) \qquad (12)$$

Note from Eq. (11) that if $\bar{T}_{eff}(E)$ is reasonably uniform over the energy range where ($f_L$ - $f_R$) is non-zero, the current will be linear with voltage and resistance will be given by (see P. 89 in Ref. [14]).

$$R^{-1} = (e^2/h)\bar{T}_{eff} \qquad (13)$$

*Dephasing model:* Phase breaking processes are described by an additional self energy $\Sigma_s$ and in/out-scattering functions $\Sigma_s^{in/out}$ ($\mp i\Sigma^{</>}$) [10,11,12,13,14]. In the first order self-consistent Born approximation, the $\Sigma_s$ and $\Sigma_s^{in/out}$ due to elastic dephasing processes are given by:

$$\Sigma_s(i,j) = \bar{D}(i,j)G(i,j) \qquad (14)$$

$$\Sigma_s^{in/out}(i,j) = \bar{D}(i,j)G^{n,p}(i,j) \qquad (15)$$

We obtain two different types of elastic dephasing by defining the matrix $\overline{D}$ as follows:

$$\overline{D}(i,j) = d_m \delta_{ij} \quad \text{("Momentum relaxing")} \quad (16)$$

$$\overline{D}(i,j) = d_p \text{ for all } i,j \quad \text{("Momentum conserving")} \quad (17)$$

where $d_m$ and $d_p$, constant factors, represent the strength of dephasing. The "momentum relaxing" dephasing relaxes both phase and momentum, while the "momentum conserving" dephasing only relaxes phase. One way to explain why the first choice for $\overline{D}$ is "momentum relaxing", while the second choice is "momentum conserving" is to note that $\overline{D}(i,j)$ can be viewed as the correlation between the dephasing potential $U_s(i)$ and $U_s(j)$ at the points '$i$' and '$j$' due to random fluctuations.

$$\overline{D}(i,j) \sim \langle U_s(i) U_s^*(j) \rangle \quad (18)$$

where $<\cdots>$ symbol represents an ensemble-average [12][14]. In a homogeneous system, $\overline{D}(i,j)$ only depends on the distance between the point '$i$' and '$j$' $(i-j)$ and not on $(i+j)$. It is well-known that the Fourier transform of the correlation of the perturbing potential at wave vector $q$ is responsible for a loss of momentum $\hbar q$ [17]. Since the choice of $\overline{D}$, independent of $(i-j)$ [Eq.(17)], has a Fourier transform of $\delta(q)$, it can not lead to any momentum loss.

$$\mathcal{F}(\overline{D}(i,j) = d_p) = \sqrt{2\pi} d_p \delta(q) \quad (19)$$

By contrast, the choice of $\overline{D}(i,j) \sim \delta_{ij}$ [Eq. (16)] should have a Fourier transform that is the same for all $q$ thus leading to momentum loss.

$$\mathcal{F}(\overline{D}(i,j) = d_m \delta_{ij}) = d_m / \sqrt{2\pi} \quad (20)$$

We will present numerical examples showing that these two choices indeed lead to the stated results. Since the most widely used method of introducing dephasing to the channel is the Buttiker probe model, one example comparing it with our dephasing models is also presented. Within the NEGF approach, we implement this dephasing model by choosing the self-energy of the Buttiker probe coupled to site '$k$' as

$$\Sigma_k(i,j) = -i\eta_k \delta_{ij} \delta_{ik} \quad (21)$$

where $\eta_k$ is zero for the site connected to the left and right contact and equal to a constant $\eta$ for the rest of sites. The electrochemical potential at each probe is then adjusted to ensure that the current drawn by each probe is zero.

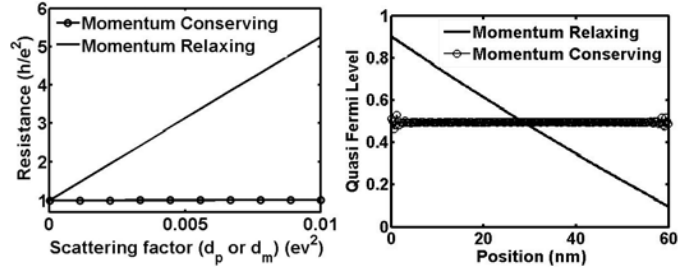

FIG. 2. (a) Resistance for a 1D conductor as a function of dephasing factor ($d_p$ or $d_m$) ($L$=60nm, $m^*$=$m_e$). (b) Quasi Fermi level of each point on the 1D conductor, normalized to '1' and '0' in the left and right contacts respectively. Actual bias used is small enough that response is linear ($d_p$=0.01ev$^2$, $d_m$=0.01ev$^2$). Circled line: with "momentum conserving" dephasing; solid line: with "momentum relaxing" dephasing.

*Results:* First we compare the effect of the "momentum relaxing" and "momentum conserving" dephasing on the device resistance [see Fig. 2(a)] for a 1D conductor (parameters provided in the caption of Fig. 2). With "momentum relaxing" dephasing, the resistance increases with increased dephasing strength [Fig. 2(a), solid line]. On the other hand, with "momentum conserving" dephasing, the resistance stays constant around the contact resistance, which is close to the quantum contact resistance ($h/e^2$) [Fig. 2(a), circled line]. For the 1D conductor, the quasi Fermi level profile across the channel for the both types of dephasing is plotted in Fig. 2(b). This is obtained from the ratio of the diagonal elements of electron correlation function $G^n(i,i)$ to the diagonal elements of the spectral function $A(i,i)$. With "momentum conserving" dephasing, the quasi Fermi level stays unchanged indicating a ballistic conductor. With "momentum relaxing" dephasing, the quasi-Fermi level changes linearly. The results in Fig. 2 show that our model for "momentum conserving" dephasing has no effect on the resistance. However, the model for "momentum relaxing" dephasing relaxes momentum and adds an additional resistance to the channel.

To compare the effect of these two types of incoherent dephasing on phase relaxation, the same 1D configuration shown in Fig. 1 is considered, but with two additional scatterers (Fig. 3, inset) whose interference leads to large oscillations in the transmission versus energy (Fig. 3, solid line). Both "momentum conserving" and "momentum relaxing" dephasing randomize phase and destroy oscillations in the transmission. Although in Fig. 3 we adjust dephasing factors ($d_p$ & $d_m$) to obtain approximately the same phase relaxation effect from both types of dephasing mechanisms, "momentum relaxing" dephasing leads to a transmission that is almost half the transmission calculated with "momentum conserving" dephasing. This reduction in transmission indicates that "momentum relaxing" dephasing adds an additional resistance to the channel. The calculated transmission with "momentum conserving" dephasing is an average of the oscillating transmission yielded from the coherent dephasing (The areas under the both curves shown with solid line and circled line in Fig. 3 are the same).

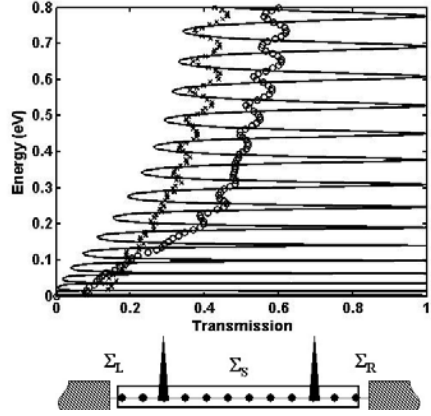

FIG. 3. Transmission as a function of energy for a 1D conductor with two scatterers as shown in the inset. Each scatterer is represented by a potential of 0.5ev at one lattice site ($L$=12nm, $m^*$=$m_e$, $d_p$=0.0012ev$^2$, $d_m$=0.012ev$^2$). Solid line: without any phase breaking processes; circled line: with "momentum conserving" dephasing; crossed line: with "momentum relaxing" dephasing.

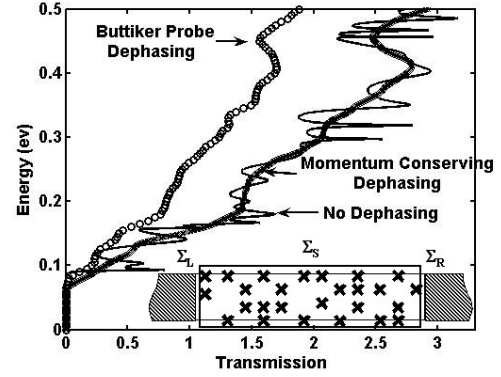

FIG. 4. Transmission as a function of energy for a 2D conductor with random scatterers (Shown in the inset). Scatterers are represented by random potentials with the random strength in range of [0eV-0.1eV]. Parameters used: $L$=25nm, $W$=10nm, $m^*$=0.2×$m_e$. Solid line: without any phase breaking processes; crossed line: with "momentum conserving" dephasing ($d_p$=0.0002ev$^2$); circled line: with Buttiker probe dephasing ($\eta$=0.01ev).

Finally, we consider a 2D quantum device with scatterers of random strength added to random sites (Fig. 4, inset), which shows sharp peaks in the transmission in the absence of any dephasing due to the quantum interference effects off coherent scatterers (Fig. 4, Solid line). Buttiker probes coupled to each site on the channel remove fluctuations in the transmission (Fig. 4, circled line), but introduces an additional resistance as evident by a reduction in the transmission. By contrast, "momentum conserving" dephasing destroys fluctuations without introducing any additional resistance (Fig. 4, crossed line). This observation confirms that "momentum conserving" dephasing only decreases phase relaxation length without any effect on mean free path.

*Summary:* In this paper, we have proposed an NEGF-based dephasing model with two specific choices of the self-energy $\Sigma_s$ that provide "momentum conserving" and "momentum relaxing" dephasing. The first one affects only phase relaxation length, while the second one affects both phase and momentum relaxation lengths. Any linear combination of these two choices can be used to adjust phase and momentum relaxation lengths independently as appropriate for a specific problem. We believe this approach provides a flexibility that is not currently available while retaining the simplicity of other phenomenological models. This work was supported by the MARCO focus center for Materials, Structure and Devices.